\documentclass[a4paper,fleqn]{cas-dc}

\usepackage[numbers,compress]{natbib}

\usepackage{subcaption}

\usepackage{amsmath}

\usepackage{wrapfig}

\usepackage[justification=centering]{caption}

\usepackage[export]{adjustbox}

\begin{document}
	
	\let\WriteBookmarks\relax
	
	\def\floatpagepagefraction{1}
	
	\def\textpagefraction{.001}
	
	
	\shorttitle{}    
	
	
	\shortauthors{}  
	
	
	\title [mode = title]{Off-resonant photoluminescence spectroscopy of high-optical quality single photon emitters in GaN}
	






\tnotetext[1]{Corresponding author}



\author[1]{Nilesh Dalla}
\cormark[1]

\affiliation[1]{organization={University of Warsaw, Faculty of Physics, Institute of Experimental Physics},
	addressline={Ludwika Pasteura 5}, 
	city={Warsaw},
	postcode={02-093}, 
	state={},
	country={Poland}}

\author[1]{Paweł Kulboka}

\author[1]{Michał Kobecki}

\author[1]{Jan Misiak}

\author[2]{Paweł Prystawko}

\author[2]{Henryk Turski}

\author[1]{Piotr Kossacki}

\author[1]{Tomasz Jakubczyk}

\affiliation[2]{organization={Institute of High-Pressure Physics "Unipress", Polish Academy of Sciences},
	addressline={Sokolowska 29/37}, 
	city={Warsaw},
	postcode={01-142}, 
	country={Poland}}








\begin{abstract} 
	In this work, we analyze the relevance of excitation parameters on the emission from single-photon emitting defect centers in GaN. 
	We investigate the absorption spectrum of different emitters by photoluminescence excitation technique at 10\,K. We report large spectral jumps (shifts up to 22\,meV) in the emitters' zero-phonon line (ZPL). The likelihood of such jumps is increased by the change in excitation energy. The shifts could indicate a large built-in dipole moment of the defects and suggest a possibility to electrically tune their ZPL  in a wide range. From the photoluminescence excitation studies, we observe that for majority of the emitters the absorption peaks exist between 2 and 2.55\,eV. The absorption peaks vary from emitter to emitter, and no universal absorption pattern is apparent. Finally, for selected emitters we observe significantly reduced spectral diffusion and instrument-limited linewidth of $138\,\mu$eV  (0.04\,nm).These findings show a new perspective for atomic defect GaN emitters as sources of coherent photons, shine new light on their energy level structure and show the possibility of tuning the ZPL, paving the way to fully harness their potential for applications in quantum technologies.
\end{abstract}






\begin{highlights}
	
	
	\item Improved spectral diffusion
	
	\item Large build-in dipole moment
	
	\item Photoluminescence excitation studies at low temperature
	
	
\end{highlights}



\begin{keywords}
	
	Quantum emitters\sep Atomic defects\sep Single photon emitters\sep 
	
\end{keywords}

\maketitle


\section{Introduction}\label{}

GaN has been known to the scientific community for its application in high-speed electronics, high breakdown voltage, and blue light-emitting diodes because of its large band gap\cite{roccaforte2020introduction}. Point defects in GaN include native isolated defects like vacancies, interstitials, and antisites, alongside intentional or unintentional impurities, as well as complexes formed by various combinations of these isolated defects. Certain defects within the GaN crystal have been observed to exhibit optical activity, emitting fluorescence across blue, green, yellow, red and infra-red bands upon optical stimulation\cite{PhysRevLett.109.267401,berhaneBrightRoomTemperature2017,ZhouTelecomGaNemitters2018}. While emission in the telecom wavelength range is highly desirable for optimal transmission in optical fibers, telecom-wavelength single-photon emitters remain scarce despite the vast number of potential host and defect combinations \cite{ferrentiIdentifyingCandidateHosts2020,zhangMaterialPlatformsDefect2020}. A recent study indicates that non-radiative processes are significantly enhanced at these low photon energies \cite{turianskyRationalDesignEfficient2024} , resulting in decreased efficiency.  

GaN is a technologically mature wide bandgap semiconductor that offers an excellent platform for quantum emitters, making it a desirable candidate for possible applications in quantum communications. It has been recently shown that GaN hosts single photon emitters in the 600-700\,nm wavelength range, which are bright even at room temperature \cite{berhaneBrightRoomTemperature2017,BerhanePhotophysicsofGaN2018} and hence easy to work with, and that these emitters feature optically addressable spin \cite{luoRoomTemperatureOptically2024}. The defects likely originate from point defects rather than dislocations\,\cite{nguyenEffectsOfMicrostructure2019}. Theoretical and experimental works suggest that particularly N$_{\text{Ga}}$V$_{\text{N}}$ can be responsible for the visible single photon emitter defects  \cite{yuanGanAsA2023,Opticaldipolestructure}, however further works are needed to fully confirm this hyptohesis. Also, the emitters observed so far even at cryogenic temperatures featured spectral diffusion over three orders of magnitudes larger than the expected lifetime-limited linewidth \cite{yuanGanAsA2023,gengUltrafastSpectralDiffusion2023}, excluding the emitters from applications where photon indistinguishability is required. 

Our study used MOCVD-grown GaN, which balances high-quality crystal and sufficiently high concentration of optically active defects. Highest-quality samples obtained with MBE have significantly smaller concentration of such defects.  Possible reason for the discrepancy between growth techniques could be much lower unintentional  carbon contamination in MBE-grown samples \cite{lyonsFirstprinciplesUnderstandingPoint2021}. This in turn suggests that the optically active centers may be associated with carbon-related defects \cite{yuanCarbonSiliconImpurity2024}.



\section{Experimental Setup}

The sample consisted of a sapphire substrate with two GaN layers ($1.5\,\mu m$ buffer layer and $3.5\,\mu m$ of controlled carbon concentration layer) grown on it. The carbon concentration in the sample was $2.3\times10^{17}/cm^3$. To estimate the emitter density, we used a room-temperature confocal microscopy setup. The setup consisted of a green 2.33\,eV (532\,nm) laser for excitation. The laser was filtered out from the detection window between 1.85 and 2.03 \,eV (610-670\,nm). We used a 100X Mittutoyo (0.7 NA) objective to focus the laser to a diffraction-limited spot well below $1\,\mu m$. A Newport NPXY200SG XY scanner (resolution of 4 nm) served to map the sample surface. A 500 mm long spectrometer in Czerny-Turner configuration, equipped with gratings up to 2400 lines/mm and a CCD camera featuring pixel size of $16\,\mu m$ was used for spectrally-resolved detection. To perform the photoluminescence excitation (PLE) studies, we used a supercontinuum light source tunable in the 1.12-3.09\,eV (400-1100\,nm) range and a Rhodamine dye laser tunable in the 2.03-2.17\,eV (570-610\,nm) range as excitation. The sample was put in a helium bath cryostat with a variable temperature insert and kept at 10\,K. 

The high refractive index of GaN decreases the setup's collection efficiency. For experiments requiring high number of photon counts we employed a cryo-compatible immersion objective \cite{jasnyFluorescenceMicroscopySuperfluid1996} in contact with the GaN sample surface to address this issue. This approach also allowed us to significantly minimize chromatic aberrations during excitation.


\section{Results}



\begin{figure}[h]

\centering
\includegraphics[width=0.8\linewidth]{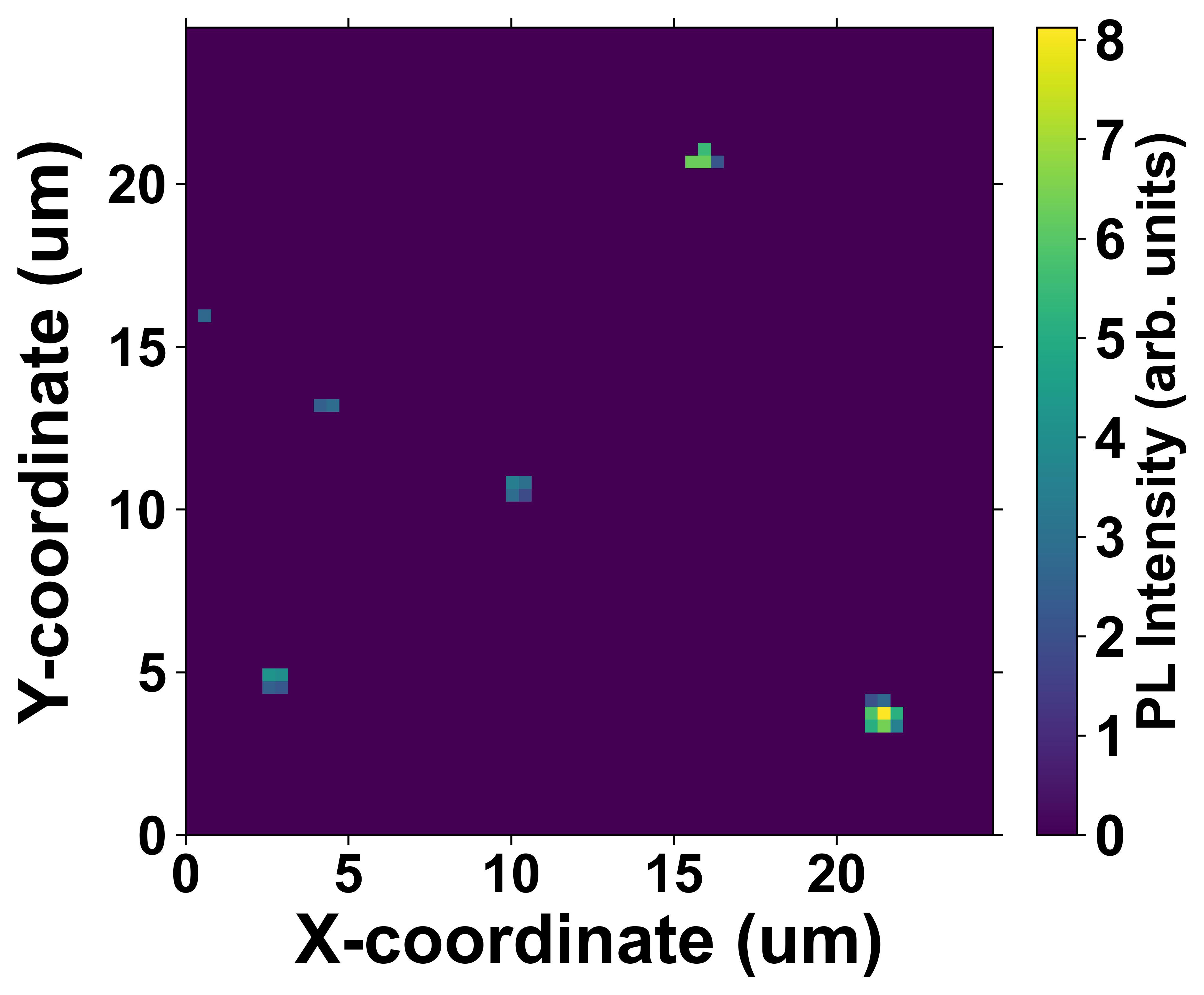}
\caption{a) Confocal scan of the sample. Bright spots correspond to emission lines in the spectral range from 1.85 to 2.03 \,eV. Each of the island of pixels corresponds to an emitter. }
\label{SampleA}


\end{figure}

To determine the concentration of defects, we performed room-temperature confocal scan maps. We focused the laser right below the surface of the GaN sample.  A typical 25x25\,$\mu m$ confocal scan is presented in Fig.\, \ref{SampleA}. Based on numerous scans, we estimate the emitter concentration to be around $0.01-0.1$ emitter/$\mu m^2$, comparable to previous findings\,\cite{berhaneBrightRoomTemperature2017}. This corresponds to volume density of $0.44-4.4\times10^{10}$ emitter/$cm^3$, assuming our focal spot has a depth of $\Delta z = 2\lambda/(n \text{NA}^2) = 2.2\,\mu m$,  where $\lambda$ is the excitation wavelength and $n$ is the GaN refractive index.  Each bright pixel corresponds to an emitter peak within the 1.85-2.03\,eV range.

%

















\begin{figure}[h]

\centering

\begin{subfigure}{0.5\textwidth}
	
	\centering
	\includegraphics[width=0.6\linewidth]{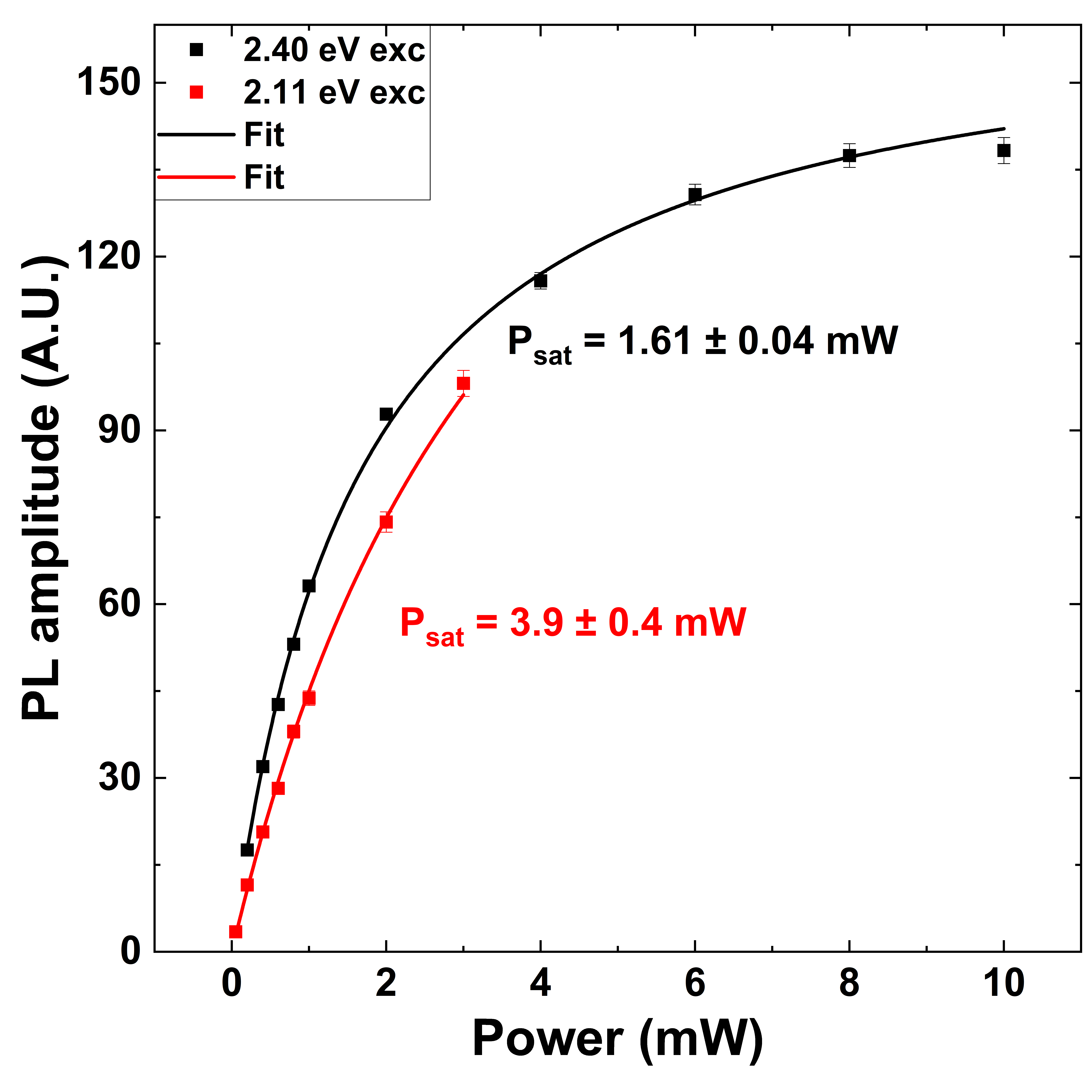}
	
	\caption{}
	
	\label{PLvspower}
	
\end{subfigure}

\begin{subfigure}{0.5\textwidth}
	\centering
	
	\includegraphics[width=0.6\linewidth]{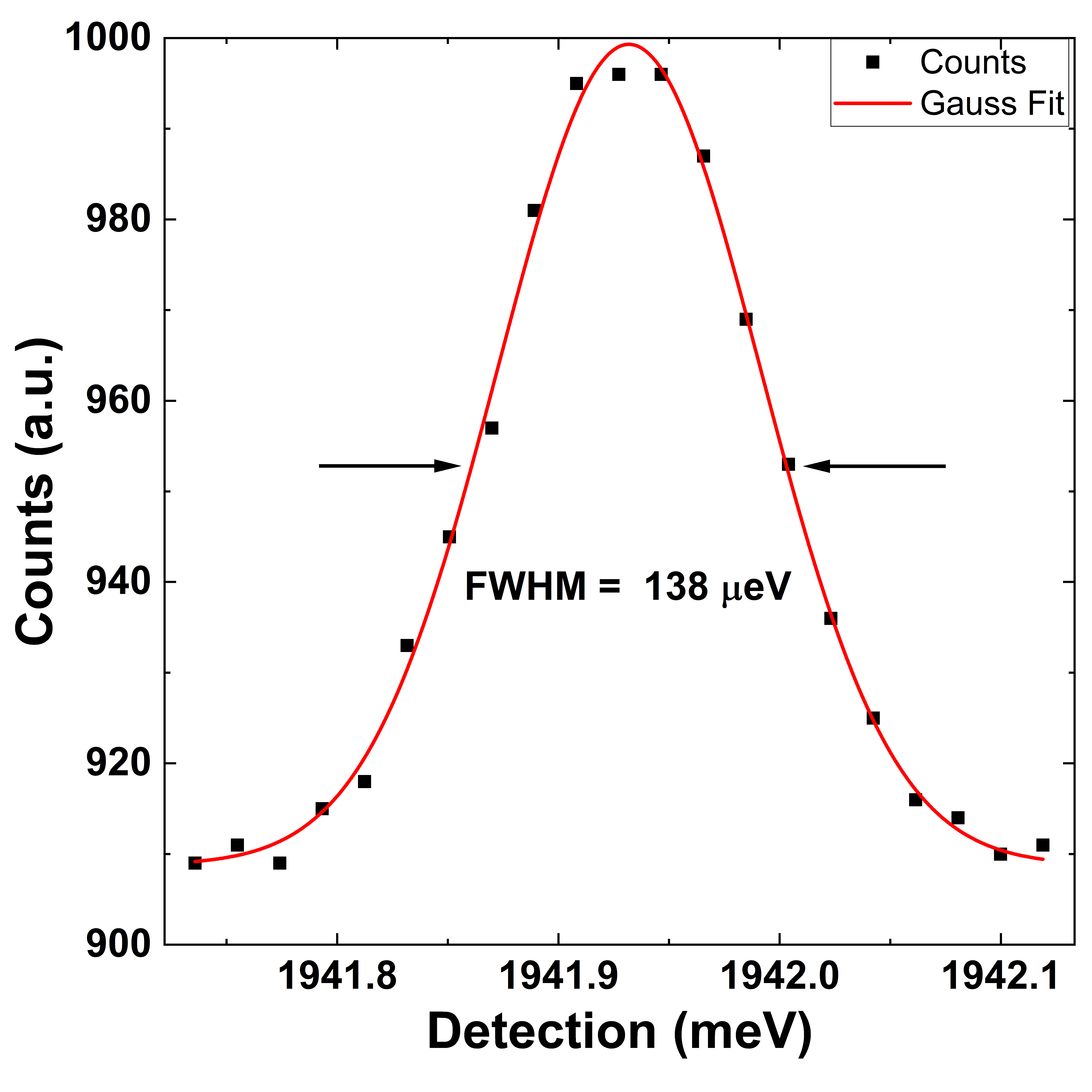}
	
	\caption{}
	
	\label{narrowline}

\end{subfigure}

\caption{a) PL amplitude vs Power curve recorded for different emitters. The red and black curves correspond to different excitation energy. 
	b)Photoluminescence spectrum of an emitter at 10\,K with 2.137\,eV excitation. Red curve corresponds to a fitted Gaussian curve with full width half maximum (FWHM) $138\,\mu$eV.}

\end{figure}

To avoid phonon-related linewidth broadening \cite{gengDephasingOpticalPhonons2023} in further PL studies, we mounted the sample on the immersion objective and cooled down the sample to 10\,K. By changing the angle of the beam entering the objective, we could selectively address different spatial locations on the sample. As previously, we focused the laser right below the surface of the GaN sample. Again, in the area accessible in this experiment (within several tens of \(\mu\mathrm{m}\) m from the optical axis), we found spatially isolated emitters with spectral lines in the range typical for the GaN emitters\,\cite{berhaneBrightRoomTemperature2017}.

First, we measured the intensity of the emission as a function of excitation power to establish optimal excitation powers as shown in Fig.\,\ref{PLvspower}. The saturation power was obtained by fitting the PL amplitude intensity with the equation 	$I=I_{0} P/(P + P_{sat})$, where $P_{sat}$ represent the saturation power and $I_{0}$ is the maximum counts number (when the population inversion would be achieved). In our experiments, the power was kept to be in the range 1-4\,mW. Please note, that due to layout of the immersion objective \cite{jasnyFluorescenceMicroscopySuperfluid1996} only up to 50\% of the laser power reaches the sample due to factors including position of the emitter with respect to the axis of objective and the size of the beam entering the pupil of the objective. The temperature evolution of the emitter line was found to be similar as reported in the literature\cite{gengDephasingOpticalPhonons2023}: the linewidth increases with temperature, and a red shift was also observed. We therefore assume, that our emitters are of the same type as the ones observed in earlier works in this spectral region. 

A noticeable difference with respect to previous reports are significantly smaller linewidths observed in our experiments.  We found several emitters featuring linewidth close to or even at the spectrometer resolution limit. An example is given in the figure \ref{narrowline}.  
Fig. \,\ref{narrowline}, displays a photoluminescence spectrum of one of the emitters at 10 K, with the excitation laser set to 2.137\,eV (580\,nm). At this temperature, the emitter line exhibits a perfect Gaussian shape, consistent with emitters previously reported in the literature. The FWHM, determined from the fit, is $138\,\mu$eV. This value corresponds to the resolution of the spectrometer, indicating that measured linewidth is constrained by our experimental setup rather than the intrinsic limit of the emitters studied.
\begin{figure}
\centering
\includegraphics[width=1\linewidth]{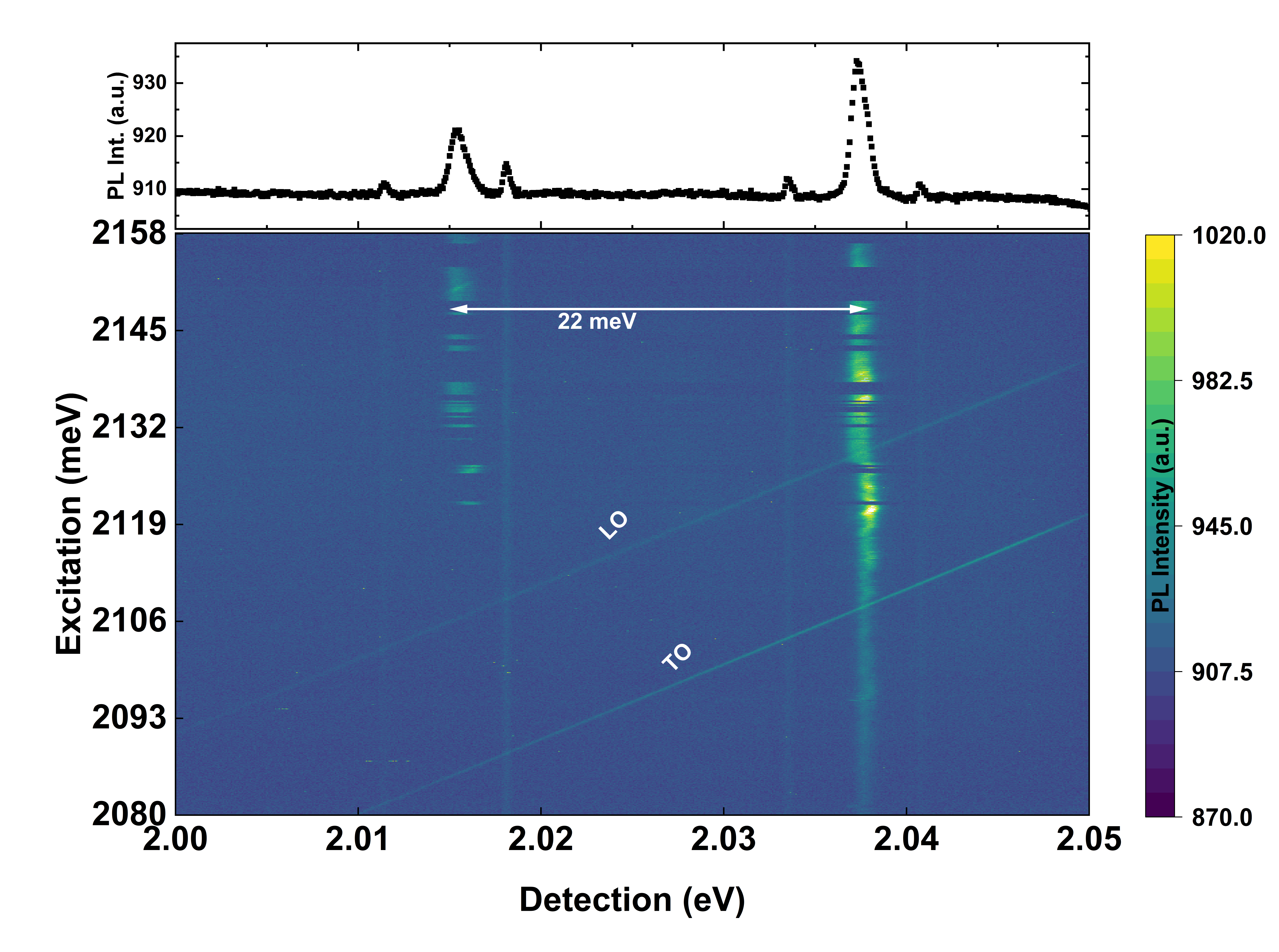}
\caption{Photoluminescence excitation map of an emitter at 10\,K with ZPL switching between two states. The black graph at top represents the averaged PL signal of the map between 2140 to 2150\,meV. Both LO and TO phonon replicas of the laser are observed. No enhancement of the PL is observed at the intersection of the replicas upon crossing with the emitter ZPL, indicating that these phonons do not play a role in the excitation mechanism.}

\label{PLE-jump}

\end{figure}


We tentatively attribute the reduction of spectral diffusion to the high quality of our MOCVD-grown material. 
Despite the improvement the ultimate lower limit of the ZPL linewidth, defined by the inverse of the lifetime of the excited state, lies two order of magnitude lower than the linewidth that we observe with the spectrometer. In previous reports\cite{gengUltrafastSpectralDiffusion2023}, as well as confirmed in our experiments lifetime of the emitter were determined to be between 1-4 ns for the emitters, corresponding to linewidth below $1\,\mu$eV.

During our experiments, we occasionally observe the emission line's giant, discrete spectral shifts, as shown in Fig.\,\ref{PLE-jump}. The change of the wavelength of the excitation laser during PLE increases the likelihood of the shifts. The shifts, however, do not deterministically depend on any controllable parameter in our experiment.  The largest observed shift was 22\,meV (see Fig.\,\ref{PLE-jump}b), much larger than shifts observed in previous works \cite{gengUltrafastSpectralDiffusion2023}. 

The large, discrete shifts could indicate the existence of isolated, single-charge traps in the vicinity of our defects—a conclusion in line with the reduced spectral diffusion observed for our defects. Numerous defect-related traps would average out to a large Gaussian spectral diffusion and fewer discrete shifts, as observed in previous reports \cite{gengUltrafastSpectralDiffusion2023,gengDephasingOpticalPhonons2023,berhaneBrightRoomTemperature2017}. 

The large magnitude of spectral shifts, not commonly observed for other solid-state defect centers, could result from Stark-effect and indicate the defects' large change in the permanent (or induced) electric dipole moment between the ground and excited states. Another mechanism that could be responsible for the spectral shifts would be the change between an ionized and neutral states of the defect. However, charge-state transitions for defects are predicted to be typically up to >1 eV \cite{yuanGanAsA2023}. The large dipole moment would also explain the vulnerability of the centers to spectral diffusion observed in earlier reports (which occurs much faster than typical integration times\,\cite{gengUltrafastSpectralDiffusion2023}). The large susceptibility could also explain the large variation in the observed ZPL energies of the defects, which covers the range from at least 1.7 to at least 2\,eV \cite{berhaneBrightRoomTemperature2017,gengDephasingOpticalPhonons2023}. 

\begin{figure}[h]

\centering

\includegraphics[width=\linewidth]{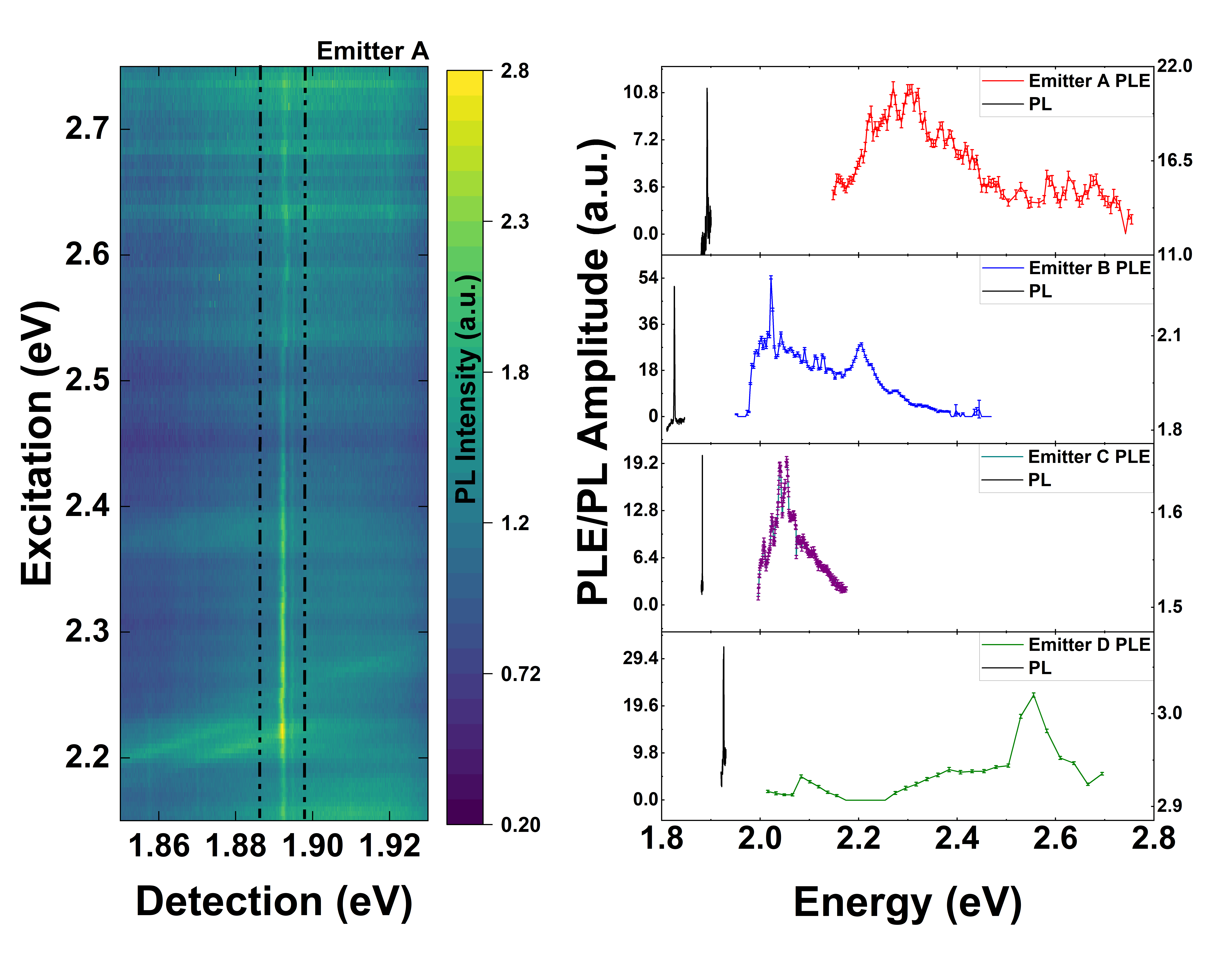}

\caption{Exemplary photoluminescence excitation spectra of different emitters. The left map represents the change in photoluminescence as the excitation is changed. On the right side, the amplitude of the PL peak is plotted against the excitation energy. The first graph (emitter A) corresponds to the map on the left-hand side. The other three graphs correspond to similar measurements performed on different emitters. The corresponding ZPL of the all the emitters is plotted with the black lines in the same graphs.}

\label{PLE-all-emitters}

\end{figure}	




%
















We performed photoluminescence excitation studies to find the optimal laser energy for excitation. The idea of the experiment is to scan the excitation energy while keeping the detection window fixed. We analyzed in this manner well above 20\, different emitters. Again, we found the emitters by scanning the laser across the sample surface. After finding an emitter line, we scanned the excitation energy by changing the energy of the tunable laser (both dye and supercontinuum lasers were used in these experiments). Figures \ref{PLE-jump} and \ref{PLE-all-emitters} show PLE maps performed in this manner at a temperature of 10\,K. In the latter case, the emitter line is present around 1.89\,eV (emitter A). The vertical dashed black lines represent the subsection of the image that was used to analyze the absorption maxima. On the right-hand side of Fig\,\ref{PLE-all-emitters} shows the amplitude of the Gaussian peak plotted with respect to the change in excitation energy for the subsection. The map presented corresponds to the emitter A as labeled. In this figure we also show the behavior of other emitters (B, C and D).  

For all the analyzed emitters, we found that a several-tens of meV broad absorption peak typically exists between around 2 and 2.55\,meV (\textasciitilde 540-620\,nm). Also, when we moved the excitation energy closer to the laser line for some emitters, we observed sharp resonances in the PLE signal (emitters B and C, for example). We tentatively attribute those features to phonon-assisted absorption. However, as it is shown in Fig. \ref{PLE-jump}, the sharp absorption lines do not match to LO or TO phonon-assisted absorption. When the energy difference between the emission line and the laser energy matches the energy of those phonons, no enhancement of the PL is observed. This suggests that the coupling to these specific bulk phonons may be weak or absent for this type of defect, possibly due to the defect’s symmetry or local vibrational modes differing from those in the GaN bulk.

\section{Conclusion}\label{}


To conclude, we observe high-quality emitters and reveal the optimal wavelength range for their excitation. We were able to find emitters with instrument-response limited linewidth ($138\,\mu$eV at 10\,K), and we performed photoluminescence spectroscopy on the emitters. 
We observe  absorption peaks at various energies and large ZPL shift of the emitters, possibly indicating their high susceptibility to electric fields.
This work contributes new insights into the photo-excitation paths and energy structure of the emitters. The findings can be used to fine-tune the excitation parameters, tuning of ZPL to coupling with micro-cavities, enabling more efficient excitation and increasing the single photon count rates of the emitters,  which is a crucial requirement for application in quantum technologies. 

\section{Acknowledgment}
This work was partially supported by the Polish National Science Centre (NCN) under Decision
 DEC-2021/43/D/ST7
 /03367. N.D., P.K., M.K. and T.J, acknowledge support from the Polish
National Agency for Academic Exchange (NAWA) under Polish Returns 2019 Programme (Grant No. PPN/PPO/2019/1/00045/U/0001). TJ acknowledge funding from the European Horizon EIC Pathfinder Open programme under grant agreement no. 101130384 (QUONDENSATE).

\break



\bibliographystyle{cas-model2-names}

\end{document}